\documentclass[showpacs,prl,superscriptaddress,nofootinbib]{revtex4}
\usepackage{graphicx}
\usepackage{color}
\usepackage{amssymb,amsfonts,amsmath}
\usepackage{natbib}
\usepackage{epstopdf}
\usepackage{subfigure}

\begin{document}

\title{Control of multidimensional systems on complex network.}

\author{Giulia Cencetti} 
\affiliation{Dipartimento di Ingegneria dell'Informazione, Universita` di Firenze,
Via S. Marta 3, 50139 Florence, Italy}
\affiliation{Dipartimento di Fisica e Astronomia and CSDC, Universit\`{a} degli Studi di Firenze, via G. Sansone 1, 50019 Sesto Fiorentino, Italia}
\affiliation{INFN Sezione di Firenze, via G. Sansone 1, 50019 Sesto Fiorentino, Italia}
\author{Franco Bagnoli}  \affiliation{Dipartimento di Fisica e Astronomia and CSDC, Universit\`{a} degli Studi di Firenze, via G. Sansone 1, 50019 Sesto Fiorentino, Italia}
\affiliation{INFN Sezione di Firenze, via G. Sansone 1, 50019 Sesto Fiorentino, Italia}
\author{Giorgio Battistelli} 
\affiliation{Dipartimento di Ingegneria dell'Informazione, Universita` di Firenze,
Via S. Marta 3, 50139 Florence, Italy}
\author{Luigi Chisci} 
\affiliation{Dipartimento di Ingegneria dell'Informazione, Universita` di Firenze,
Via S. Marta 3, 50139 Florence, Italy}
\author{Duccio Fanelli} \affiliation{Dipartimento di Fisica e Astronomia and CSDC, Universit\`{a} degli Studi di Firenze, via G. Sansone 1, 50019 Sesto Fiorentino, Italia}
\affiliation{INFN Sezione di Firenze, via G. Sansone 1, 50019 Sesto Fiorentino, Italia}

\begin{abstract}
Multidimensional systems coupled via complex networks are widespread in nature and thus frequently invoked for a large plethora of interesting applications.
From ecology to physics, individual entities in mutual interactions are grouped in families, homogeneous in kind. These latter interact selectively, through a sequence of self-consistently regulated steps, whose deeply rooted architecture is stored in the assigned matrix of connections. The asymptotic equilibrium eventually attained by the system, and its associated stability, can be assessed by employing standard nonlinear dynamics tools. For many practical applications, it is however important to externally drive the system towards a desired equilibrium, which is resilient, hence stable, to external perturbations. To this end we here consider 
a system made up of $N$ interacting populations which evolve according to general rate equations, bearing attributes of universality. One species is added to the pool of interacting families and used as a dynamical controller to induce novel stable equilibria. Use can be made of the root locus method to shape the needed control, 
in terms of intrinsic reactivity and adopted protocol of injection.  The proposed method is tested on both synthetic and real data, thus enabling to demonstrate its robustness and  versatility. 
\end{abstract}


\maketitle

Investigating the interlinked dynamics of an ensemble composed of units organized in homologous families, constitutes a universal challenge in science, of broad and cross-disciplinary breath~\cite{Murray03, CrossHohenberg93, Pikovsky03}. Each population is customarily identified in terms of its continuous density. This latter evolves in time, as dictated by specific self-reaction stimuli, that generally bear nonlinear contributions.  In a complex and dynamical environment, species experience a large plethora of mutual interactions,  declinated via different modalities, notably  pairwise exchanges. Cooperative and competitive interference are simultaneously at play, and shape the ultimate fate of the system as a whole~\cite{SuweisSiminiBanavarMaritan13}. These fundamental ingredients,  flexibly combined and properly integrated, are at the roots of any plausible mathematical model targeted to community interactions~\cite{CaldarelliChessa16}, from ecology~\cite{CoyteSchluterFoster15} to neuroscience~\cite{KandeSchwartzJessell00,AsllaniChallengerPavoneSacconiFanelli14}, passing from genetic and human health~\cite{Lodishetal00}, through a full load of man-made technological applications~\cite{RohdenSorgeTimmeWitthaut12}. Irrespectively of the specific realm of investigation, each population can be abstractly assigned to a given node of a virtual graph. Directed or indirected edges among nodes exemplify the topological structure of the existing network of interactions~\cite{Boccaletti_etal14, Boccaletti_etal06}. The intricate web of inter-species connections, key information to anticipate the expected dynamics of the system, is therefore encoded in the associated adjacency matrix~\cite{BarratVespignani08, ArenasDiazGuileraKurthsMorenoZhoug08}. 
 
In many cases of interest, it is important to drive the system towards a desired equilibrium, that is stable, and thus resilient, to external perturbations~\cite{GaoBarzelBarabasi16, GrilliRogersAllesina16, ThebaultFontaine10, DorflerBullo12, Cencetti_etal17}. For example, hostile pathogens could be forced to go extinct: the stability of the attained equilibrium would efficaciously shield from subsequent harmful invasion and outbreaks. Alternatively, it could prove vital to robustly enhance the expression of species identified as beneficial for the system at hand. Building on these premises, we here develop and test a general control strategy~\cite{Kalman63, Luenberger79, SlotineLi91, LiuSlotineBarabasi11, NicosiaCriadoRomanceRussoLatora11} targeted to multidimensional systems consisting of a large number of components that interact through a complex network. By inserting one additional species, the controller, which configures as a further node of the collection, we will be able to manipulate the asymptotic dynamics of the system, in terms of existence and stability of the allowed fixed points.  

To set the reference frame we will hereafter consider a system consisting of $N$ species  (nodes) whose activities $\boldsymbol{x}=(x_1,x_2, \cdots, x_N)^T$ obey the coupled nonlinear equations~\cite{GaoBarzelBarabasi16, CoyteSchluterFoster15, GrilliRogersAllesina16}:

\begin{equation}
  \dot x_i=f_i(x_i) + \sum_jA_{ij} g_i(x_i,x_j) \ \ \ \ i=1,\dots,N.
  \label{init}
 \end{equation}

The first term on the right-hand side specifies the self-dynamics of species $i$ while the second term stems from the interactions of species $i$ with the other species. The nonlinear functions $f_i(x_i)$ and $g_i(x_i, x_j)$ encode the dynamical laws that govern the system's components, while the weighted connectivity matrix $\boldsymbol{A}$ captures the interactions between nodes. The elements $A_{ij}$ can be positive or negative, depending on the specific nature of the interaction, i.e. cooperative or competitive. Notice that system~\eqref{init} is assumed in~\cite{GaoBarzelBarabasi16} as a reference model to analyze resilience patterns in complex networks. Differently from~\cite{GaoBarzelBarabasi16}, $A_{ij}$ can here take positive and negative values (see also~\cite{TuGrilliSuweis16}).

In ecological applications, the number of nodes reflects the biodiversity of the scrutinized sample~\cite{May72, CoyteSchluterFoster15, GrilliRogersAllesina16}. Distinct trophic layers materialize as coherent blocks in the adjacency matrix, whose entries modulate the strength of mutual interactions~\cite{SuweisSiminiBanavarMaritan13}. These are often epitomized by a quadratic response function $g_i(x_i, x_j)$~\cite{GrilliRogersAllesina16}.  Each species is then subjected to a reaction drive $f_i(x_i)$, usually a logistic growth with a prescribed carrying capacity~\cite{GaoBarzelBarabasi16}. Animals displaying competitive predator-prey interactions or, alternatively,  subjected to a symbiotic dependence, such as in plant-pollinator relationships, are among the systems that fall within the aforementioned scenario~\cite{SuweisSiminiBanavarMaritan13}. Furthermore, the complex community of micro-organisms that live in the digestive tracts of humans and other animals, including insects, can be rooted on similar descriptive grounds~\cite{CoyteSchluterFoster15}. For genetic regulatory networks, the dynamical variables $x_i$ represent the level of activity of a gene or the concentration of the associated proteins~\cite{BecskeiSerrano00}. Species specific reaction terms $f_i(x_i)$ account for e.g. degradation or dimerization. The pattern of activation could be effectively modeled by sigmoidal Hill-like functions~\cite{Murray03}, as follows the classical Michaelis-Menten scheme~\cite{JohnsonGoody11}, which incorporates the known map of gene interactions. On a more general perspective, understanding the emerging dynamics in social communities~\cite{Cavallaro_etal14}, grasping the essence of the learning organization in the brain~\cite{Nicosiaetal13}, and implementing efficient protocols for robot navigation in networked swarms~\cite{Rubenstein_etal14} are among the very many applications that can be traced back to one of the variants of equations~\eqref{init}, with a suitable choice of the nonlinear functions $f_i(x_i)$ and $g_i(x_i, x_j)$.  

\vspace{1 truecm}
{\bf Adding a species to enforce stable equilibria in a multidimensional system}\\

Starting from the above illustrated setting, we will here discuss a suitable control scheme to drive system~\eqref{init} towards a desired equilibrium  $\boldsymbol{x}^*=(x_1^*,x_2^*, \cdots, x_N^*)^T$, which is linearly stable to externally imposed perturbations. To reach this goal we shall introduce one additional species, the $(N+1)$-th component of the collection, suitably designed to yield the sought effect. To set the notation, we indicate by $u$ the component (e.g., concentration, activation level) assigned to the controller and write:  

 \begin{equation}
 \left\{\begin {array}{ll} 
  \dot x_i=f_i(x_i) + \sum_jA_{ij} g_i(x_i,x_j) + \alpha_i h_i(x_i,u)\\
  \dot u=  - (u-u^*)- \rho \sum_j\beta_j(x_j-x_j^*).\\
\end{array}\right.
 \label{added_node}
 \end{equation}

The controller $u$ can exert a direct influence on every component $x_i$, as  specified by newly added terms $\alpha_i h_i(x_i,u)$ that modify the original  system~\eqref{init}. 
$\boldsymbol{\alpha}=(\alpha_1,\alpha_2,...,\alpha_N)^T$ is a vector of $N$ constant parameters, to be self-consistently adjusted following the scheme depicted below.  $h_i(x_i,u)$ is a generic, in principle nonlinear, function of the components $x_i$ and $u$ that reflects the modality of interactions between the controller and the existing species. The equation for the dynamical evolution of the controller $u$ displays two distinct contributions. 
The first represents a self-reaction term, assumed to be linear just for ease of presentation. The nonlinear self-dynamics of the controller $u$ can be readily considered, with no further technical complication.
The rate of change of $u$ is assumed to be contextually driven by a global forcing that senses the relative distance of $x_i$ from its deputed equilibrium $x_i^*$.  The parameters $\boldsymbol{\beta}=(\beta_1,\beta_2,...,\beta_N)^T$ and $\rho$ will prove central in enforcing the stabilization of the prescribed fixed point. A few comments are mandatory to fully appreciate the generality of the  proposed framework, beyond the specific choices made for purely demonstrative purposes. Let us begin by remarking that the controller $u$ can represent an artificially engineered component or, equivalently, belong to an extended pool of interacting populations. 
In the scheme here imagined, it is assumed that the values of $u$ and $x_i$ $\forall i$,  are accessible to direct measurements at any time and that this information can be processed to set the controller dynamics. This is largely reasonable for experiments that run under protected conditions like, e.g., the study of microbial dynamics in laboratory reactors, but certainly less realistic for applications that aim at in vivo multidimensional systems, think for instance to genetic regulatory circuits. The dynamical equation for $u$ can, however, be amended  to a large extent and with a great deal of flexibility, depending on the target application and the structural specificity of the employed controller,  while still allowing for an analogous methodological treatment\footnote{As a matter of fact, we can equivalently assume a generalized equation for the controller of the type $\dot u=f_u(u)-\rho g_u(\boldsymbol x,u,\boldsymbol\beta)+b$ where $f_u(u^*)=0$ and $b=\rho g_u(\boldsymbol x^*,u^*,\boldsymbol\beta)$.}. The dynamics of the original, unsolicited, components and the functional form that specifies the controller feedback bear unequivocal universality traits~\cite{GaoBarzelBarabasi16}. 

The global fixed point ($\boldsymbol{x}^*,u^*$) of the controlled system~\eqref{added_node} should match the following constraints

 \begin{equation}
  f_i(x_i^*) + \sum_jA_{ij} g_i(x_i^*,x_j^*) + \alpha_i h_i(x_i^*,u^*)=0 \quad i=1,..,N
  \label{fixed_point}
 \end{equation}
 which, provided the $x_i^*$ and $u^*$ are assigned, ultimately set the values of the parameters $\alpha_i$. Conversely, as we shall illustrate in the following, one could assume the parameters $\alpha_i$ as a priori known and infer via equations~\eqref{fixed_point} the fixed point(s) to be eventually stabilized. The next step in the analysis aims at ensuring the stability of the selected fixed point. This will be achieved by acting on the residual free parameters $\boldsymbol{\beta}$ and $\rho$. As routinely done, we perturb the equilibrium solution as $x_i=x_i^*+v_i$, $u=u^*+w$ and Taylor expand equations~\eqref{added_node} assuming the imposed disturbances $\boldsymbol\eta=(\boldsymbol{v},w)$ small in magnitude. At the linear order of approximation one obtains:
\begin{equation}
 \boldsymbol{\dot\eta}=
 \left (
  \begin{matrix}
  \boldsymbol{G} & \boldsymbol{q}\\
 -\rho \boldsymbol\beta^T & -1\\
 \end{matrix}
 \right )\boldsymbol{\eta} \equiv \boldsymbol{J} \boldsymbol{\eta}
 \end{equation}
 where $\boldsymbol q$ is a $N$-dimensional column vector of components $q_i=\alpha_i\frac{\partial h_i}{\partial u}(x_i^*,u^*)$. The $N\times N$ matrix  $\boldsymbol G$ is defined as: 
 
\begin{eqnarray*}
 G_{ii}&=&\frac{\partial f_i}{\partial x_i}(x_i^*) + \sum_kA_{ik}\frac{\partial g_i}{\partial x_i}(x_i^*,x_k^*) + \alpha_i\frac{\partial h}{\partial x_i}(x_i^*,u^*) \\
G_{ij} &=& A_{ij}\frac{\partial g_i}{\partial x_j}(x_i^*,x_j^*).\\
 \end{eqnarray*}
 
The fixed point ($\boldsymbol{x}^*,u^*$) is linearly stable if all eigenvalues of the Jacobian matrix $\boldsymbol{J}$ have negative real parts. The associated characteristic polynomial $P(\lambda)=\det(\boldsymbol J-\lambda\boldsymbol I)$ can be cast in the equivalent, affine in the $\rho$-parameter, form:  
 
 \begin{eqnarray*}
 &&P(\lambda) = (-1)^{N+1}(1 + \lambda)\det(\boldsymbol{G}-\lambda\boldsymbol{I})-\rho\sum_{i,j=1}^N\beta_j [\text{adj}(\boldsymbol{G}-\lambda\boldsymbol{I})]_{ji}q_i \equiv \mathcal D(\lambda) + \rho \mathcal N(\lambda) 
 \label{char_eq}
 \end{eqnarray*}
 that is reminescent of the celebrated root locus method~\cite{Evans48}.  Here, $[\text{adj}(\boldsymbol{Z})]_{ji}=(-1)^{i+j}\det[(\boldsymbol{Z})_{(i,j)}]$ denotes the adjugate of matrix $\boldsymbol Z$, $( \boldsymbol{Z} )_{(i,j)}$ being the minor of $\boldsymbol{Z}$ obtained by removing the $i$-th row and the $j$-th column. The polynomials $\mathcal D(\lambda)=-\prod^{N+1}_{k=1}(\lambda-p_k),$ and $\mathcal N(\lambda)=\prod^{N-1}_{k=1}(\lambda-z_k)$ have respectively degrees $N+1$ and $N-1$.
With a slight abuse of language we will refer to as {\it poles} the roots $p_k$ of the polynomial $\mathcal D(\lambda)$ and {\it zeros} the roots $z_k$ of $\mathcal N(\lambda)$. Notice that for $\rho=0$ the eigenvalues of the Jacobian $\boldsymbol{J}$  correspond to the $N+1$ poles $p_k$. These latter quantities are uniquely determined, once the fixed point ($\boldsymbol{x}^*,u^*$)  has been assigned. In particular it cannot a priori be ensured that the real parts of all $p_k$ are negative, as stability would require. In other words, when $\rho=0$, we can enforce the desired fixed point into the system but cannot guarantee its stability. On the other hand, for $\rho \rightarrow \pm \infty$, $N-1$ eigenvalues of $\boldsymbol{J}$ tend to the zeros $z_k$, which depend self-consistently on the free parameters $\boldsymbol\beta$. As we shall show hereafter, it is in principle possible to assign the $\beta_i$ to force the real parts of all $z_k$ to be negative.  The two remaining eigenvalues of  matrix $\boldsymbol{J}$, in the limit of large $|\rho|$,  diverge to infinity in the complex plane. More precisely, they travel along opposite directions following  a vertical (resp. horizontal) asymptote, if $\rho$ is bound to the positive (resp. negative) semiaxis. To confer stability in the limiting case $\rho\rightarrow\infty$ where $N-1$ eigenvalues of $\boldsymbol{J}$ coincide with the roots of $\mathcal N(\lambda)$, it is therefore sufficient to (i) operate a supervised choice of $\boldsymbol\beta$ and (ii) impose the condition that yields a vertical asymptote ($\rho \rightarrow +\infty$), while, at the same time, requiring that this latter intersects the negative side of the real axis. In this respect, it is important to remark that the intersection occurs in the point of abscissa  $\lambda_0=\frac{1}{2}\bigl(\sum^{N+1}_{k=1}p_k-\sum^{N-1}_{k=1}z_k\bigr)$. Hence, the idea is to interpolate between the two limiting cases $\rho=0$ and $\rho\rightarrow\infty$ by determining the minimal value $\rho_c$ of $\rho$ beyond which the desired fixed point becomes stable. The existence of the threshold $\rho_c$ that makes the imposed fixed point attractive for any $\rho>\rho_c$ is obvious, being stability already assured in the limiting setting $\rho \rightarrow +\infty$\footnote{In principle, more than one value of $\rho_c$ can exist for which the eigenvalues cross the imaginary axis, making stable an unstable fixed point. The intersections are found imposing $\lambda=i\omega$ in equation $\mathcal D(\lambda) + \rho_c \mathcal N(\lambda)=0$, which yields a system of two equations, for respectively the real and imaginary parts. This system can then be solved for the two unknowns $\rho_c$ and $\omega$.}. For the sake of clarity we reiterate 
 that this amounts to selecting $Re(z_k)<0$ for all $k$ and further imposing $\lambda_0<0$, by properly assigning the free parameters $\boldsymbol\beta$.

In order to study the assignability of the zeros $z_k$ by means of $\boldsymbol\beta$, let us recall that for a generic square matrix $\boldsymbol Z$,
  $\text{adj}(\boldsymbol{Z}-\lambda\boldsymbol{I})=-\sum_{m=0}^{N-1}\sum_{l=0}^{N-m-1}c_{l+m+2}\boldsymbol{Z}^m\lambda^l$
 where $c_k$ stands for the coefficients of the characteristic polynomial of $\boldsymbol Z$, namely
 $\det(\boldsymbol{Z}-t\boldsymbol{I})=\sum_{l=0}^{N}c_{l+1}t^l$.
The polynomial $\mathcal N(\lambda)$ can be consequently written as:
 \begin{equation}
 \mathcal N(\lambda)=\sum_{m=0}^{N-1}\sum_{l=0}^{N-m-1}c_{l+m+2}[\boldsymbol{\beta}^T\boldsymbol{G}^m\boldsymbol{q}]\lambda^l\equiv \sum_{n=0}^{N-1}d_{n+1}\lambda^n
  \label{def_d}
 \end{equation}
 It is hence straightforward to establish a direct relation between the parameters $\boldsymbol\beta$ and the vector of coefficients $\boldsymbol d$:
 \begin{equation}
  d_n=\sum_{k=0}^{N-n}c_{k+n+1}[\boldsymbol{\beta}^T\boldsymbol{G}^k\boldsymbol{q}]
 \end{equation}
 that can also be equivalently stated as:
 \begin{equation}
  \boldsymbol{d}=\boldsymbol{H}\boldsymbol\beta
  \label{dHb}
 \end{equation}
 where $\boldsymbol H$ is the matrix defined by:
 \begin{equation}
  H_{nm}=\sum_{k=0}^{N-n}c_{k+n+1}(\boldsymbol{G}^k\boldsymbol{q})_m.
  \label{H}
 \end{equation}

The suited vector $\boldsymbol\beta$ is thus obtained\footnote{For obvious consistency reasons $\boldsymbol\beta$ must have real entries. This follows naturally if one chooses the zeros $z_k$ to be real or complex conjugate in pairs, which implies that the coefficients $d_n$ of the polynomial $\mathcal N(\lambda)$ (see~\eqref{def_d}) are real. All other quantities involved are real by definition.} from~\eqref{dHb}, provided matrix $\boldsymbol H$ is invertible. This latter request defines the condition of {\it controllability} for the scheme that we have implemented (see Supplementary Information, SI, for a discussion that aims at positioning this observation in the context of standard control theory~\cite{Kailah80}). Summing up, the devised strategy consists of the following steps. First, the fixed point is selected 
and the parameters $\boldsymbol\alpha$ frozen to their respective values as specified by Eq.~\eqref{fixed_point}. Then the complex roots $z_k$ are chosen so that $Re(z_k)<0$ for all $k$ while, at the same time, matching the condition that makes the vertical asymptote cross the horizontal axis with a negative intercept. As we will clarify when discussing the applications, 
the $z_k$ can be chosen to coincide with the poles $p_k$, except for punctual modifications whenever $Re(p_k)>0$. Notice however that $z_k$ should be real or come in conjugate pairs, as the coefficients  $d_k$ are, by definition, real.  Once the roots $z_k$ have been fixed, one can readily compute the associated polynomial coefficients
$d_k$, and hence proceed with the determination of $\boldsymbol\beta$ via~\eqref{dHb}, provided that the controllability condition holds. Finally, by selecting $\rho > \rho_c >0$ we obtain a linearly stable fixed point ($\boldsymbol{x}^*,u^*$) for the controlled dynamics~\eqref{added_node}.

\vspace{1 truecm}
{\bf Testing the control method: from synthetic gene network to real microbiota dataset}\\

\begin{figure}
 \centering
 \subfigure[]
   {\includegraphics[width=6cm]{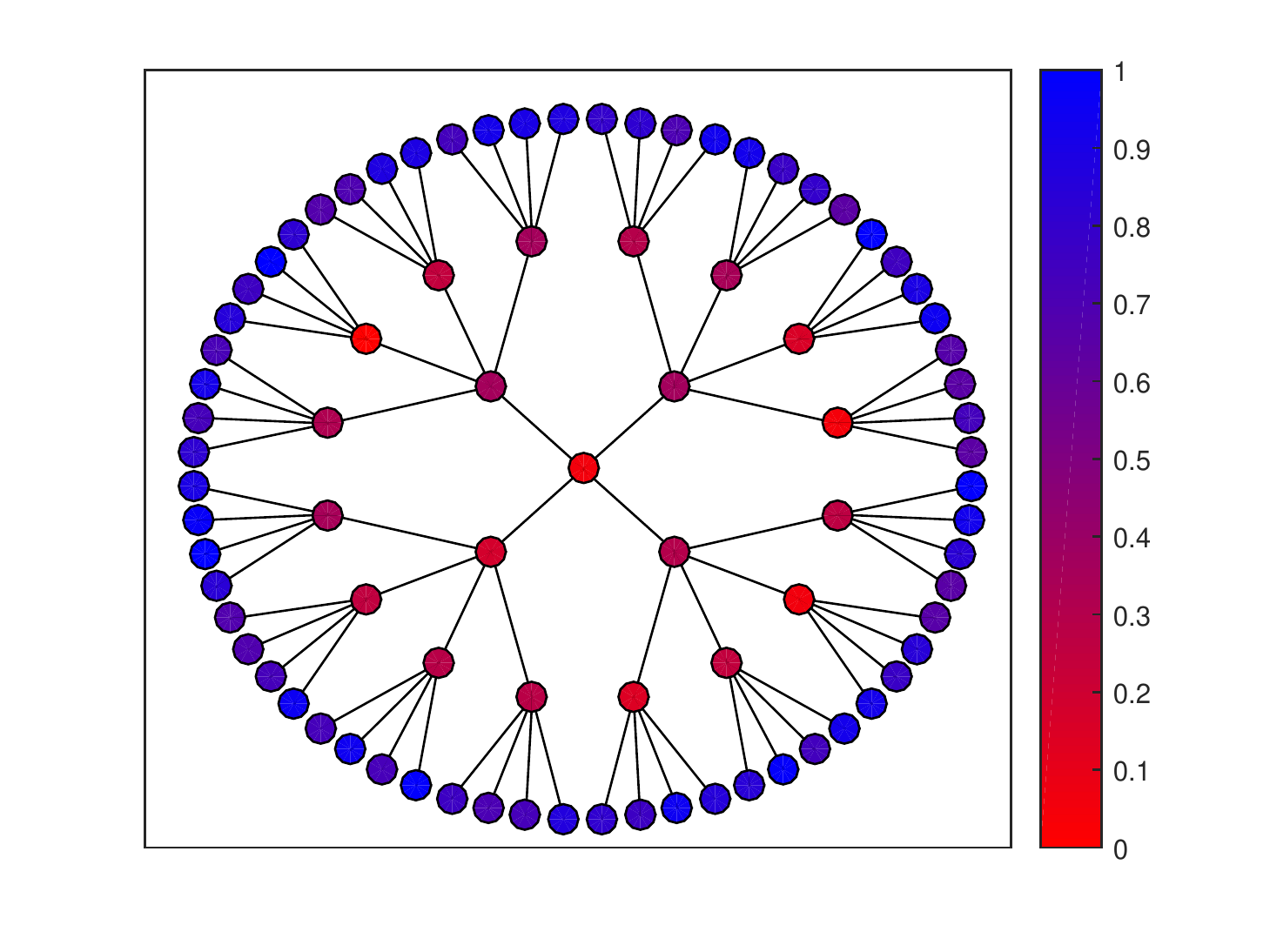}}
 \hspace{5mm}
  \centering
 \subfigure[]
   {\includegraphics[width=6cm]{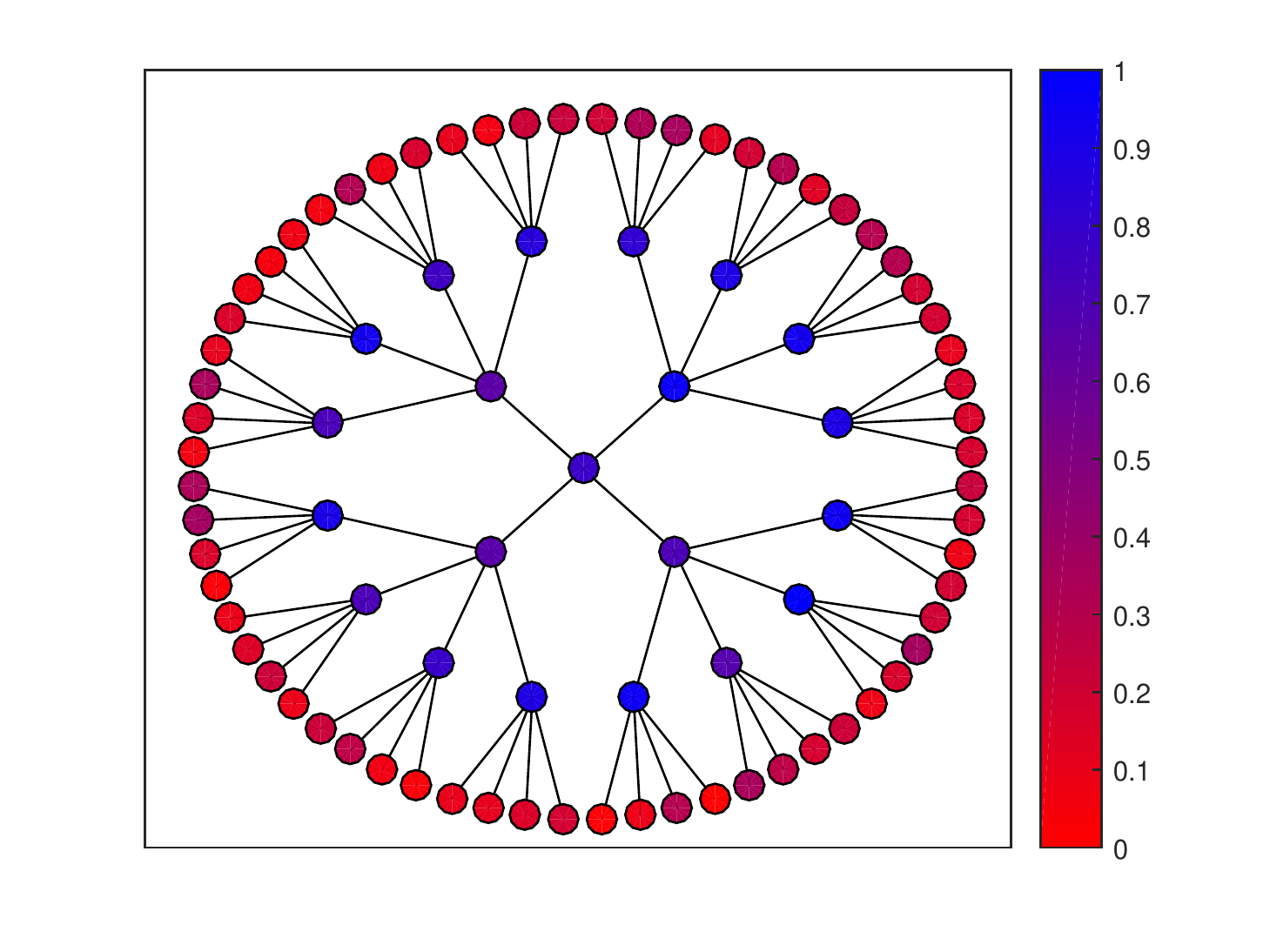}}
 \hspace{5mm} 
  \centering
 \subfigure[]
   {\includegraphics[width=15cm]{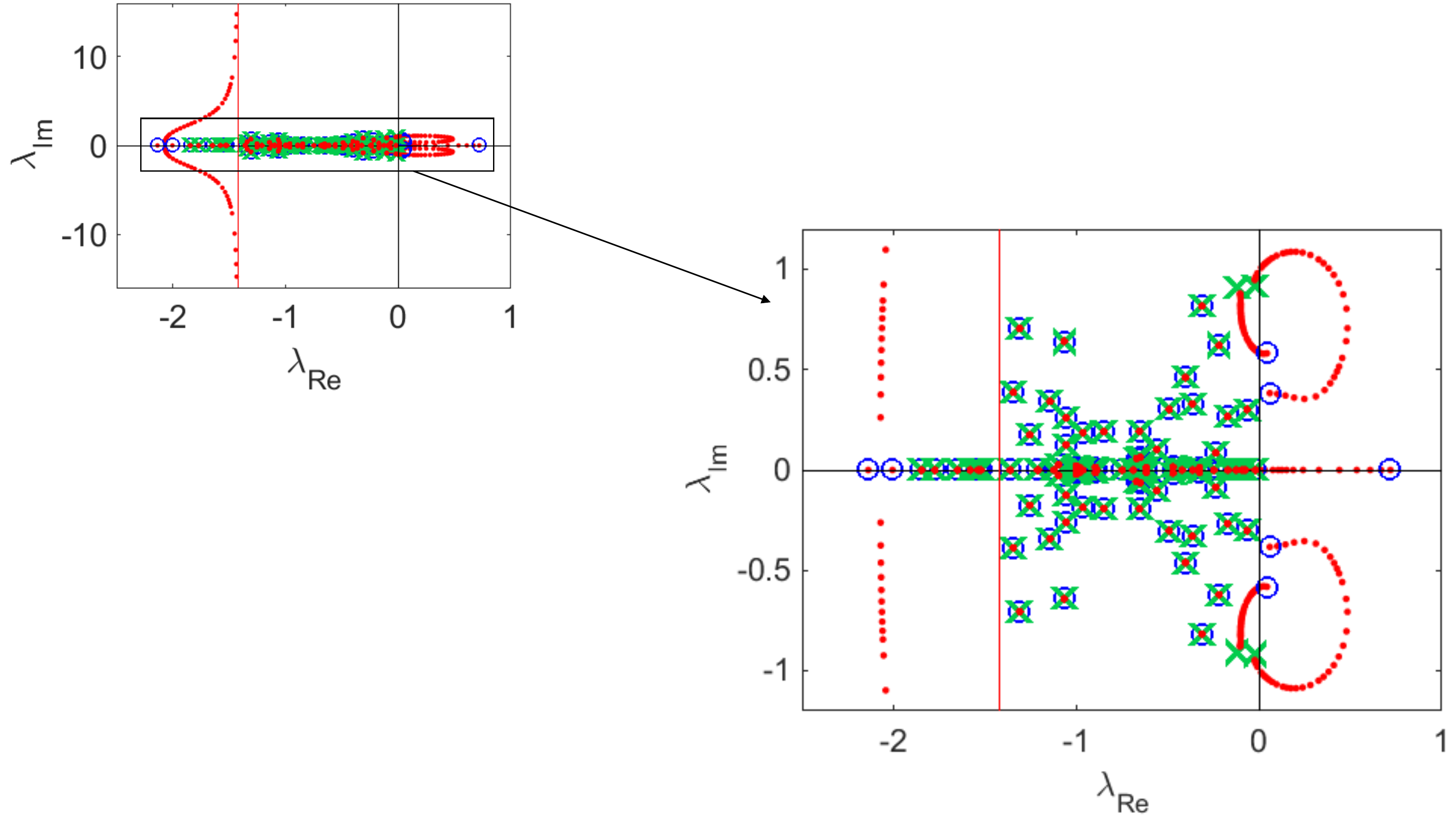}}
 \hspace{5mm}
   \caption{Panel (a): the control is modulated so as to enhance the activity of the peripheral nodes of the tree, as compared to the inner ones. Panel (b): the control makes now the bulk nodes more active as compared to the peripheral ones. Panel (c):  the root locus diagram relative to the situation displayed in panel (b) is plotted. Blue circles stand for the position of the complex eigenvalues when $\rho=0$, while green crosses identify the eigenvalue obtained for $\rho \rightarrow \infty$. The vertical red line represents the asymptote that attracts two of the modified eigenvalues, when $\rho \rightarrow \infty$. The red dots show the computed spectrum, calculated when increasing $\rho$. In this case the matrix  $\boldsymbol{A}$ contains an identical number of $\pm 1$ entries. These are randomly assigned and kept unchanged for all tests performed. The figure on the right is a zoom of the plot displayed on the left.}
   \label{fig1}
  \end{figure}

As a first application of the above technique, we will study the dynamics of an artificial gene network~\cite{AlbertRooman11, HastyMcMillenIsaacsCollins01, ElowitzLeibler00, IsaacsHastyCantorCollins03}. In our example the network of connections is a regular tree with branching ratio
$r = 4$. It is further assumed that the genetic activation between nodes $i$ and $j$ is described in terms of a Hill function, with  cooperation coefficient equal to $2$. In formulae, $A_{ij}=1$ and 
$g_i(x_i,x_j) \equiv g(x_j) = x_j^2/(1+x_j^2)$. Negative regulation loops are also accommodated for. These latter could, in principle, be modeled by assuming paired interactions of the type $1-g(x_j)$, while still setting to one the relative entry of the connection matrix. As described in the SI, we can equivalently set $A_{ij}=-1$, while assuming interactions to be modulated by $g(x_j)$ as indicated above. At the same time, the reaction part should be modified with an additional term, $\eta_i$, counting the number of negative loops that affects node $i$. More specifically,  $f(x_i) = - \gamma_i x_i +\eta_i$, where the first term mimics constitutive degradation. In our tests, matrix $\boldsymbol{A}$ contains an identical number of randomly assigned $\pm1$. The parameters $\gamma_i$ are random variables uniformly distributed over the interval $[0,1]$.  Working in this setting our aim is to control the equilibrum state of the system and thus shape the pattern of asymptotic activity. For this initial application we choose to operate with a simple linear control and set  $h_i(x_i,u) \equiv h(u) = u$, for all $i$. In this case, $u$ could e.g. represent the density of a suitable retroviral vector used to infect specific cell lines~\cite{SignaroldiLaise_etal16}.
To provide an immediate graphical illustration of the power of the method, we set to stabilize two distinct fixed points. In the first example, see Figure~\ref{fig1}(a), the control is designated so as to enhance the degree of activity of the peripheral nodes of the tree.  These latter are characterized by a similar value of the activity, apart for slight randomly superposed fluctuations. Similarly, the nodes that define the bulk of the tree 
display a shared degree (except for tiny stochastic  modulation) of residual activity. In Figure~\ref{fig1}(b), the dual pattern is  instead obtained and stabilized: the peripheral nodes are now being silenced and the activity concerns the nodes that fall in the center of the tree. In Figure~\ref{fig1}(c) the root locus diagram relative to the situation reported in Figure~\ref{fig1}(b) is displayed. By properly tuning $\rho$ above a critical threshold $\rho_c$, we can enforce the stability of the obtained fixed point. Two eigenvalues diverge to $\pm \infty$ following a vertical asymptote in the complex plane. For each chosen fixed point that is being stabilized the zeros $z_k$ can be selected so as to make the asymptote intercept the horizontal axis in the left-half of the plane.

As a second application of the proposed control strategy, we set to study the dynamics of the gut microbiota~\cite{CoyteSchluterFoster15}. 
The intestinal microbiota is a microbial ecosystem of paramount importance to human health~\cite{Shenetal15}.  Efforts are currently aimed at 
understanding the microbiota ability to resist to enteric pathogens and assess the response to antibiotics 
cure of intestinal infections. Recent advances in DNA sequencing and metagenomics
make it possible to quantitatively characterize the networks of interactions that rule the dynamics of the 
microbiota ecosystem. This was for instance achieved in~\cite{SteinBuccietal13} by analyzing available data on mice~\cite{BuffieJarchumetal12} with an innovative approach which combines classical Lotka-Volterra model 
and regression techniques. Eleven species were identified and thoroughly analyzed in terms of self and mutual dynamics. 

In the following we shall apply the method here developed to control the dynamics of the whole microbioma~\cite{SteinBuccietal13} or a limited sub-portion of it.  In this specific case, the self-dynamics is assumed to be  logistic, namely $f_i(x_i)=x_i(r_i-s_ix_i)$, while $g(x_i,x_j)=x_i x_j$. The constants $r_i$ and $s_i$ are provided in~\cite{SteinBuccietal13} and follow from direct measurements.  The weighted matrix of connections  $\boldsymbol{A}$ 
presents both positive and negative entries, assigned according to~\cite{SteinBuccietal13}.  Finally, $h_i(x_i,u)=u x_i$. The results of the analysis are organized under different headings that reflect the three distinct control strategies explored.

\vspace{1 truecm}

\begin{figure}
  \includegraphics[width=17cm]{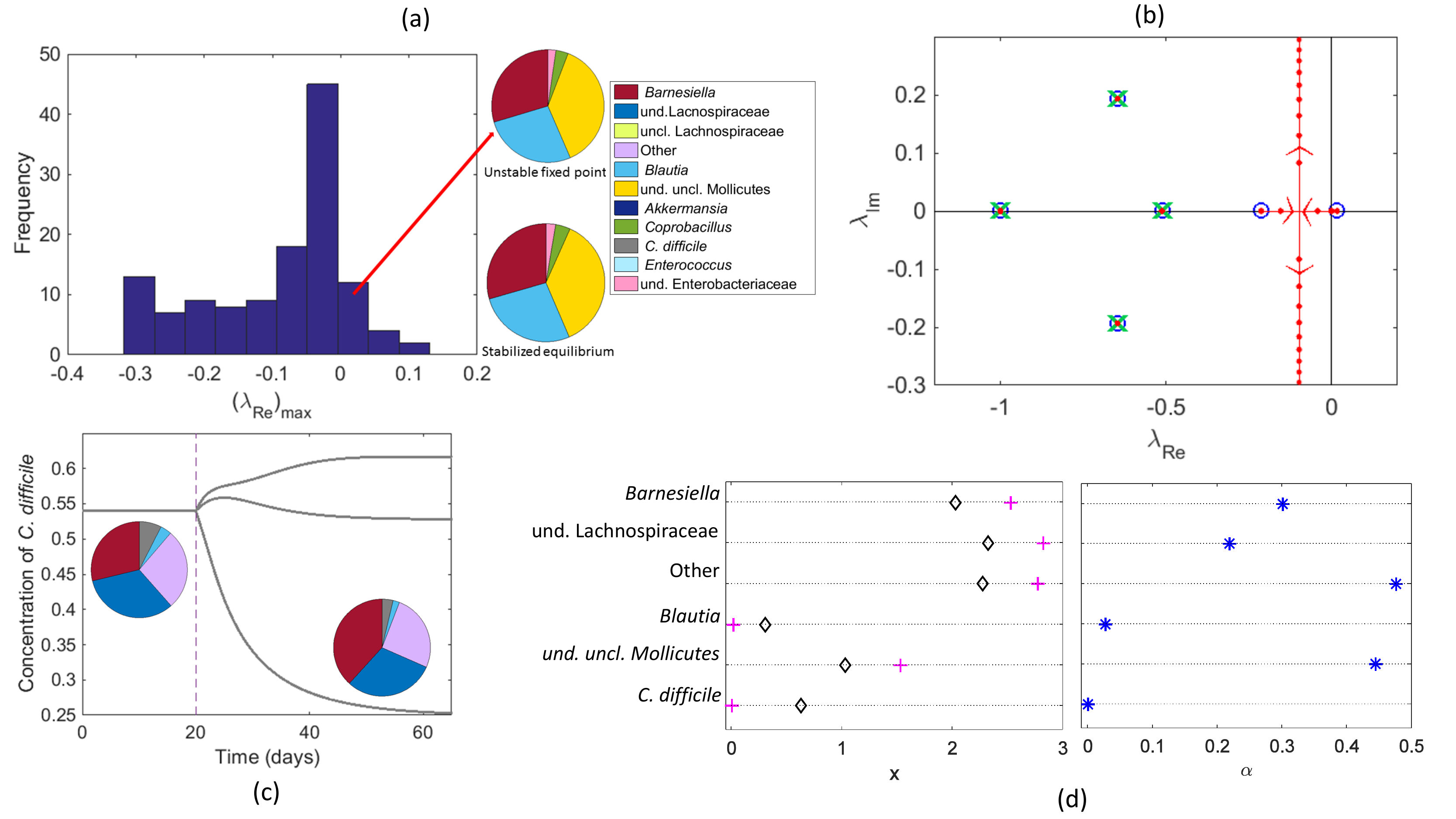}
 \caption{Panel (a): a reduced 5 species subsystem of the microbiota is considered (case A) and all possible fixed points computed. Only those displaying non-negative concentrations are retained and their stability assessed. 
   In the main figure,  the histogram of $(\lambda_{Re})_{max}$, the largest real parts  of the eigenvalues obtained after the linear stability analysis, is depicted. The two pie charts refer to the initially unstable fixed point (upper chart) and the stabilized equilibrium (lower chart). Panel (b): the root loci diagram relative to the case discussed in panel (a), is shown. Blue circles identify the position of the complex eigenvalues when $\rho=0$, while green crosses stand for the eigenvalues obtained in the limit $\rho \rightarrow \infty$. The vertical red line is the asymptote that eventually attracts the two residual eigenvalues. The red dots show the computed spectrum, when progressively increasing $\rho$. Panel (c): the goal is here to reduce the concentration of the pathogen species,  {\it C. difficile}, by employing as controller one of the species that compose the microbioma (case B). The concentration of {\it C. difficile} is monitored over time for three different control strategies, turning on the control at the same time ($t=20$ days). The insertion of the species of uncl. Lachnospiraceae provoques a substantial reduction (50 $\%$) of the pathogen concentration, as also displayed by the enclosed pie charts (for interpreting the color-code refer to panel (a)). Panel (d): we now modify
 a stable fixed point, by driving to extinction one of the existing population, the pathogen {\it C. difficile} (here species $6$), with an indirect control strategy (case C). 
 The obtained concentrations are reported in the left graph (pluses) and confronted with the initial unperturbed solution (diamonds). As anticipated $x_6^* \simeq 0$. The components of  $\boldsymbol\alpha$ are plotted in the right graph. Notice in particular that $\alpha_6=0$. }
  \label{fig2}
 \end{figure}

\textbf{\textit{Stabilizing an unstable fixed point by means of an external controller (Case A).}}
Consider the system of $11$ species, as defined in~\cite{SteinBuccietal13} (see SI for a discussion on the bacterial species involved). For illustrative purposes, we will restrict the analysis to all sub-systems that combine $5$ out of the $11$ species analyzed in~\cite{SteinBuccietal13}. The fixed points for the obtained 5 species systems are calculated. Those displaying positive concentrations are then retained for subsequent analysis.
The stability of each selected fixed point is established upon evaluation of the spectrum of the Jacobian of the reduced dynamics. In Figure~\ref{fig2}(a) the histogram of $(\lambda_{Re})_{max}$, the largest real parts of the recorded eigenvalues, is plotted: several fixed points exist that correspond to unstable equilibria. Starting from this setting, we will introduce a suitably shaped controller, following the above discussed guidelines, in order to stabilize a slightly perturbed version of an originally unstable fixed point, see pie charts in Figure~\ref{fig2}(a). Denote by $\boldsymbol{x^*}$ the fixed point to be eventually stabilized and consequently assign the parameters $\alpha_i$ so as to match Eqs.~\eqref{fixed_point}. The spectrum of the Jacobian matrix obtained for $\rho=0$ (blue circles in Figure~\ref{fig2}) protrudes into the right  half-plane. More specifically, 
one eigenvalue exhibits a positive real part, so flagging the instability that one aims to control.  At variance, the crosses in Figure~\ref{fig2}(b)  stand for the roots $z_k$ of $\mathcal N(\lambda)$, and fall in the left side of the complex plane. The vertical (red, in Figure~\ref{fig2}(b)) line identifies the  location of the two residual eigenvalues of the Jacobian matrix, when $\rho \rightarrow \infty$. By tuning the parameter $\rho$, one can continuously bridge the two above limiting settings, as graphically illustrated in Figure~\ref{fig2}(b).  When $\rho>\rho_c \simeq 0.01$, the eigenvalues populate the left half-hand plane and stability is, therefore, gained. 

\vspace{1 truecm}

\textbf{\textit{Acting with one species of the pool to damp the concentration of the pathogens (Case B).}}
Select now a stable fixed point, mixture of five distinct species. One of them is {\it Clostridium difficile},  a species of Gram-positive spore-forming bacteria that may opportunistically dominate the gut flora, 
as an adverse effect of antibiotic therapy. As controller we shall here employ one of the other $6$ species that compose the microbioma\cite{Steinway_etal15, Freilich_etal11}. The aim is to drive the system towards another equilibrium, stable to linear perturbations, which displays a decreased pathogen concentration. In this case the parameters $\boldsymbol\alpha$ are determined a priori, once the control species has been identified. Denote by $\boldsymbol{\bar{A}}$ the reduced $5 \times 5$ matrix that specifies all paired interactions between the pool of populations involved in the initial fixed point. The equilibrium solution that can be attained by the controlled system is determined as  $\boldsymbol x^*=-\boldsymbol{\bar{A}}^{-1}(\boldsymbol r+\boldsymbol\alpha)$, and clearly depends on the species used as controller.  The only meaningful solutions are those displaying non negative components $x_i^*$. In the example depicted in Figure~\ref{fig2}(c) only three solutions can be retained, namely the ones obtained by using uncl. Lachnospiraceae, uncl. Mollicutes and \textit{Enterococcus} as respective control. In one of the inspected cases,  the amount of {\it C. difficile} is found to reduce, when the control is turned on. The asymptotic concentration that is eventually attained is sensibly lower than the one initially displayed. The pie charts in Figure~\ref{fig2}(c)  represent, respectively, the initial fixed point and the final stationary equilibrium, as shaped by the control in the most beneficial case, i.e., when the 
concentration of {\it C. difficile}  is seen to shrink. The root locus plot obtained for this specific case is reported in the SI. Importantly, the discussed scheme can be straightforwardly modified so as to account for a generic nonlinear self-reaction dynamics for the control species, e.g., a logistic growth,  that could replace the linear Hookean-like term assumed in Eq.~\eqref{added_node}. 

\vspace{1 truecm}

\textbf{\textit{Driving to extinction one species, the other being the target of the control (Case C).}}
As an additional example, we wish to modify a stable fixed point of the dynamics, by silencing one of the existing populations with an indirect control. In other words we shall introduce and stabilize a novel fixed point, that displays a negligible residual  concentration of the undesired species, by acting on the other species of the collection. This is for instance relevant  when aiming at, e.g., eradicating a harmful infection that proves resistant to direct therapy. With this in mind, we consider a reduced  ecosystem consisting of $6$ species, selected among the $11$ that define the microbiota. A stable fixed point exists (black diamonds in Figure~\ref{fig2}(d), left panel) which displays a significant concentration of {\it C. difficile}, the pathogen species. Assign to this latter species the index $6$. We now insert a controller which cannot directly interfere with {\it C. difficile}. This amounts, in turn, to setting to zero the corresponding component of vector $\boldsymbol\alpha$ ($\alpha_6=0$). We then require the concentration of the {\it C. difficile} to be small, i.e., $x_6^*=\varepsilon<<1$. This latter condition translates into a constraint that should be matched by the other $5$ species, namely  $\sum_{j\neq6}\bar{A}_{6j}x_j^* =   (s_6- \bar{A}_{66})\varepsilon -r_6$. Given $x_k^*$, the components $\alpha_k$, with $k \ne 6$, are chosen so as to match the constraint $\alpha_k = ( -r_k+s_kx_k^* - \sum_{j\neq6}\bar{A}_{kj}x_j^* - \bar{A}_{k6}\varepsilon)/u^*$. A possible solution of the problem is reported in Figure~\ref{fig2}(d): in the left panel (plus symbols) the components of the fixed point stabilized by the control are shown. As anticipated, the concentration of {\it C. difficile} is small. The right panel of Figure~\ref{fig2}(d) shows the components of the vector $\boldsymbol\alpha$ that specify the characteristics of the introduced controller. Notice that $\alpha_6=0$ so that the controller is not directly influencing the rate of production of {\it C. difficile}.

\vspace{1 truecm}
{\bf Discussion}\\

We would like to draw the attention on the interpretation of $\boldsymbol\alpha$. As stated earlier,  $\boldsymbol\alpha$ characterizes the strength of the coupling between the controller $u$ and every single species of the system to be controlled. An alternative interpretation is however possible: $u$  could represent a mixture of different species and the components of  
$\boldsymbol\alpha$ incorporate the relative abundance of the mixed compounds. In light of the above, also the previously discussed control schemes which apparently assumed dealing with an artificially designed control, could be realized via a proper mixture of exisiting microbiota species so as to achieve the coupling $\boldsymbol\alpha$ corresponding to the desired fixed point.\\
Notice also that the control scheme here developed could be in principle exploited to drive the system towards a stable fixed point of the unperturbed dynamics, starting from out-of-equilibrium initial conditions. To achieve this goal $u^*$ needs to be set to zero, thus requiring that the controller is turned off at equilibrium. In this case, $\boldsymbol{\alpha}$ and $\boldsymbol{\beta}$ are not subjected to specific constraints, as the existence and stability of the desired equilibrium are a priori granted. Such parameters could hence be chosen so as to reflect the specificity of the target system. In the annexed SI we demonstrate this intriguing possibility.

Summing up, we have here proposed and tested a method to control the dynamics of  multidimensional systems on a complex graph. The original system is made up of $N$ interacting populations obeying
a set of general equations, which bear attributes of universality. One additional species, here referred to as the controller, is inserted and made interact with the existing constellation of species. By tuning the strength of the couplings (or equivalently the composition of the inserted controller), we can drive the system towards a desired equilibrium. The stability of the achieved solution is enforced by adjusting the parameters that ultimately govern the rate of change of the controller. Methodologically, we make use of the root locus method which can be naturally invoked once the control problem is suitably formulated. 
The tests that we have performed, both synthetic and drawn from real life applications, demonstrate the versatility and robustness of the proposed scheme. This latter configures therefore as a viable and innovative tool to tackle a large plethora of inter-disciplinary systems, from life science to man-made applications, that should be stably driven towards a  desired configuration.  In this current implementation, and for purely pedagogical reasons, the control assumes that the state of the system is accessible to direct measurement. Relaxing this working hypothesis is a possibility that we shall explore in a future contribution.

 \vspace{1 truecm}
 \begin{section}{Supplementary Information}

{\bf Genetic network model}\\

We shall here justify the model of  genetic regulatory network analyzed in the main text. Consider first a small regulatory  network consisting of one gene (whose activity is labelled $x$) and one protein (associated to the continuous concentration $y$). A positive regulation loop can be modeled as:
 
\begin{eqnarray*}
\dot{x} &=& k_1 g(y)-\gamma_1 x \\ 
\dot{y} &=& k_2 x-\gamma_2 y 
\end{eqnarray*}
where:

\begin{equation}
g(y)=\frac{y^n}{K+y^n}.
\stepcounter{equation}
\tag{S\theequation}
\end{equation}

In the following, we will make the choice $n=2$ and $K=1$. Similarly, a negative regulation loop can be modeled as:

\begin{eqnarray*}
\dot{x} &=& k_1 (1-g(y))-\gamma_1 x \\ 
\dot{y} &=& k_2 x-\gamma_2 y 
\end{eqnarray*}

As it should be, the concentration of proteins grows with the level of gene activity. As a first step approximation, and to avoid dealing with two distinct families of mutually interlinked constituents (genes and proteins), we can replace $y$ with $x$ in the argument of $g(\cdot)$, via adiabatic elimination (apart from redefinition of the involved constants). Building on the above we model the extended regulatory network as:

\begin{equation}
\dot{x}_i = k_i \sum_j A_{ij} g(x_j)-\gamma_i x_i + k_i \eta_i 
\stepcounter{equation}
\tag{S\theequation}
\end{equation}
where $x_i$ stands for the activity of gene $i$.  
For {\it positive} feedbacks between species $i$ and $j$, $A_{ij} = 1$,  while, for {\it negative} loops $A_{ij} = -1$.
The parameter $\eta_i$ stands for the number of negative loops (number of negative entries of the $i$-th row of $\boldsymbol A$) that are associated to node $i$. Finally, to keep the structure as simple as possible we set  
$k_i=1$ for all $i$. The matrix $\boldsymbol A$ employed in the example reported in the main body has a simple structure and it has been chosen for purely illustrative purposes: it represents a regular tree network with branching ratio $r=4$.

 \vspace{1 truecm}
{\bf Controlling the Microbiota network: the experimental data}\\

 The second application of the proposed control method deals with a model of gut microbiota. As explained in the main body of the paper, the model is well established and builds on a generalization of the celebrated Lotka-Volterra equations:
\begin{equation}
  \dot x_i=x_i(r_i-s_ix_i) + x_i\sum_{j\ne i}A_{ij} x_j \ \ \ \ i=1,\dots,N
  \stepcounter{equation}
\tag{S\theequation}
  \label{SI_init}
 \end{equation}
The above equations take into account species-species interactions and self-regulation, this latter effect being described in terms of  a logistic growth. Despite its intrinsic simplicity, the model is often invoked to 
explain how ecological interactions, e.g., mutualism and competition for nutrients, can lead to complex phenomena, as multi-stability or antibiotic mediated catastrophic losses of biodiversity. In~\cite{SteinBuccietal13} the model has been applied to a relatively small (mice gut) microbiota system made up of $11$ distinct populations. More precisely, the ten most abundant species have been identified: all together they account for the vast majority ($\sim90\%$) of the total populations found in the mice gut. The remaining populations are grouped into a unique (non-homogeneous) category referred to as ``Other''.
The authors of~\cite{SteinBuccietal13} analyzed the experiments reported in~\cite{BuffieJarchumetal12} and provided a quantitative characterization of the coefficients that enter the definition of the relevant quantities $\boldsymbol r$, $\boldsymbol s$ and $\boldsymbol A$. These latter are reported in table~\eqref{tab_dataset} together with the names of the involved species:
 
 \begin{equation}
 \centering
  \resizebox{\textwidth}{!}{ 
 \begin{tabular}{c | c | c | c c c c c c c c c c c}
  Populations & $\boldsymbol r$ & $\boldsymbol s$ & & & & & & $\boldsymbol A$  \\ \hline 
  {\it Barnesiella} & 0.368 & 0.205 & 0 & 0.0984 & 0.167 & -0.165 & -0.143 & 0.0199 & -0.515 & -0.392 & 0.346 & 0.00888 & -0.269\\
  undefined genus of Lachnospiraceae & 0.310 & 0.105 & 0.0621& 0 & -0.0430 & -0.155 & -0.187 & 0.0270 & -0.459 & -0.414 & 0.301 & 0.0221 & -0.196\\
  unclassified Lachnospiraceae & 0.356 & 0.102 & 0.144 & -0.192 & 0 & -0.140 & -0.165 & 0.0136 & -0.504 & -0.772 & 0.292 & -0.00596 & -0.206\\
  Other & 0.540 & 0.831 & 0.224 & 0.138 & 0.000459 & 0 & -0.224 & 0.220 & -0.205 & -1.01 & 0.666 & -0.0390 & -0.400\\
  {\it Blautia} & 0.709 & 0.709 & -0.180 & -0.0513 & -5,03$\times10^{-5}$ & -0.0542 & 0 & 0.0162 & -0.507 & 0.554 & 0.157 & 0.224 & 0.10635\\
  undefined genus of unclassified Mollicutes & 0.471 & 0.423 & -0.111 & -0.037 & -0.0426 & 0.0410 & 0.261 & 0 & -0.185 & -0.4326 & 0.165 & -0.0610 & -0.265\\
  {\it Akkermansia} & 0.230 & 1.21 & -0.127 & -0.186 & -0.122 & 0.381 & 0.400 & -0.161 & 0 & 1.390 & -0.379 & 0.192 & -0.0963\\
  {\it Coprobacillus} & 0.830 & 4.35 & -0.0712 & 6.04$\times10^{-4}$ & 0.0803 & -0.455 & -0.503 & 0.169 & -0.562 & 0 & 0.443 & -0.223 & -0.207\\
  {\it Clostridium difficile} & 0.392 & 0.0558 & -0.0375 & -0,0333 & -0.0499 & -0.0904 & -0.102 & 0.0323 & -0.182 & -0.303 & 0 & 0.0144 & -0.00767\\
  {\it Enterococcus} & 0.291 & 0.192 & -0.0422 & -0.0131 & 0.0240 & -0.118 & -0.329 & 0.0207 & 0.0548 & -2.10 & 0.111 & 0 & 0.0238\\
  undefined genus of Enterobacteriaceae & 0.324 & 0.384 & -0.374 & 0.278 & 0.249 & -0.168 & 0.0840 & 0.0337 & -0.232 & -0.395 & 0.314 & -0.0388 & 0\\
 \end{tabular}
 }
 \stepcounter{equation}
\tag{S\theequation}
 \label{tab_dataset}
 \end{equation}
 
 The concentration of the species are measured in $10^{11}$ rRNAcopies/cm$^3$, the coefficients $\boldsymbol r$ have the dimension of the inverse of time (measured in days), and $\boldsymbol s$ is expressed as the inverse of the product of a time for a concentration.
 
The second application discussed in the main body of the paper, i.e. that targeted to controlling the microbiota dynamics, assumes the above parameters. Notice that the set of considered species includes the spore-forming pathogen {\it Clostridium difficile}. To lower its concentration (and so diminish the probability of infection) is one of the goals of the implemented control. As explained in the paper, three different control schemes are considered. In the following we will provide some additional information for each of the analyzed schemes. The physical dimension of the inserted controller $u$ is again $10^{11}$ rRNAcopies/cm$^3$. The parameters  $\boldsymbol \beta$ have dimension of the inverse of a time, while $\rho$ is a-dimensional.

 \vspace{1 truecm}
\textbf{\textit{Case A: Stabilizing an unstable fixed point by means of an external controller.}}
   The first of the three different control strategies for the microbiota network discussed in the main body involves only $5$ out of the $11$ populations being examined (more precisely {\it Barnesiella}, {\it Blautia},  und. Mollicutes, {\it Coprobacillus} and  und. Enterobacteriaceae). An unstable fixed point of equation~\eqref{SI_init}, which contains only these latter species, has been calculated and denoted by $\bar{\boldsymbol{x}}$ (see table~\eqref{tab_s1}). The maximum eigenvalue of the Jacobian matrix evaluated at $\bar{\boldsymbol{x}}$  is $(\lambda_{Re})_{max}=0.0148>0$, thus implying instability. As discussed in the main body of the paper, it is possible to stabilize a different equilibrium, denoted by $\boldsymbol x^*$ in~\eqref{tab_s1}, which corresponds to a slight modification of the unstable solution $\bar{\boldsymbol{x}}$. Following the procedure detailed in the paper one can readily calculate the parameters $\boldsymbol\alpha$ and $\boldsymbol\beta$~\eqref{tab_s1}. The values obtained are reported  
 in  table~\eqref{tab_s1}. Direct simulations of the controlled system, as displayed in Figure~\ref{fig_SI_1}, confirms that stability has been indeed achieved.  
 
  \begin{equation}
 \begin{tabular}{c | c | c | c | c}
  Populations & $\bar{\boldsymbol{x}}$ & $\boldsymbol x^*$ & $\boldsymbol\alpha$ & $\boldsymbol\beta$ $\times10^4$ \\ \hline 
  {\it Barnesiella} & 0.9736 & 0.9917 & 0.0186 & 0.1860\\
  {\it Blautia} & 0.8840 & 0.9093 & 0.0089 & -0.0638\\
  und. uncl. Mollicutes & 1.2361 & 1.2396 & 0.0089 & -0.0032\\
  {\it Coprobacillus} & 0.1169 & 0.1363 & 0.1005 & 0.0495\\
  und. Enterobacteriaceae & 0.0756 & 0.0894 & 0.0175 & -2.2067\\
  \label{tab_s1}
 \end{tabular}
 \stepcounter{equation}
\tag{S\theequation}
 \end{equation}
 
\begin{figure}
\centering
 \includegraphics[width=11cm]{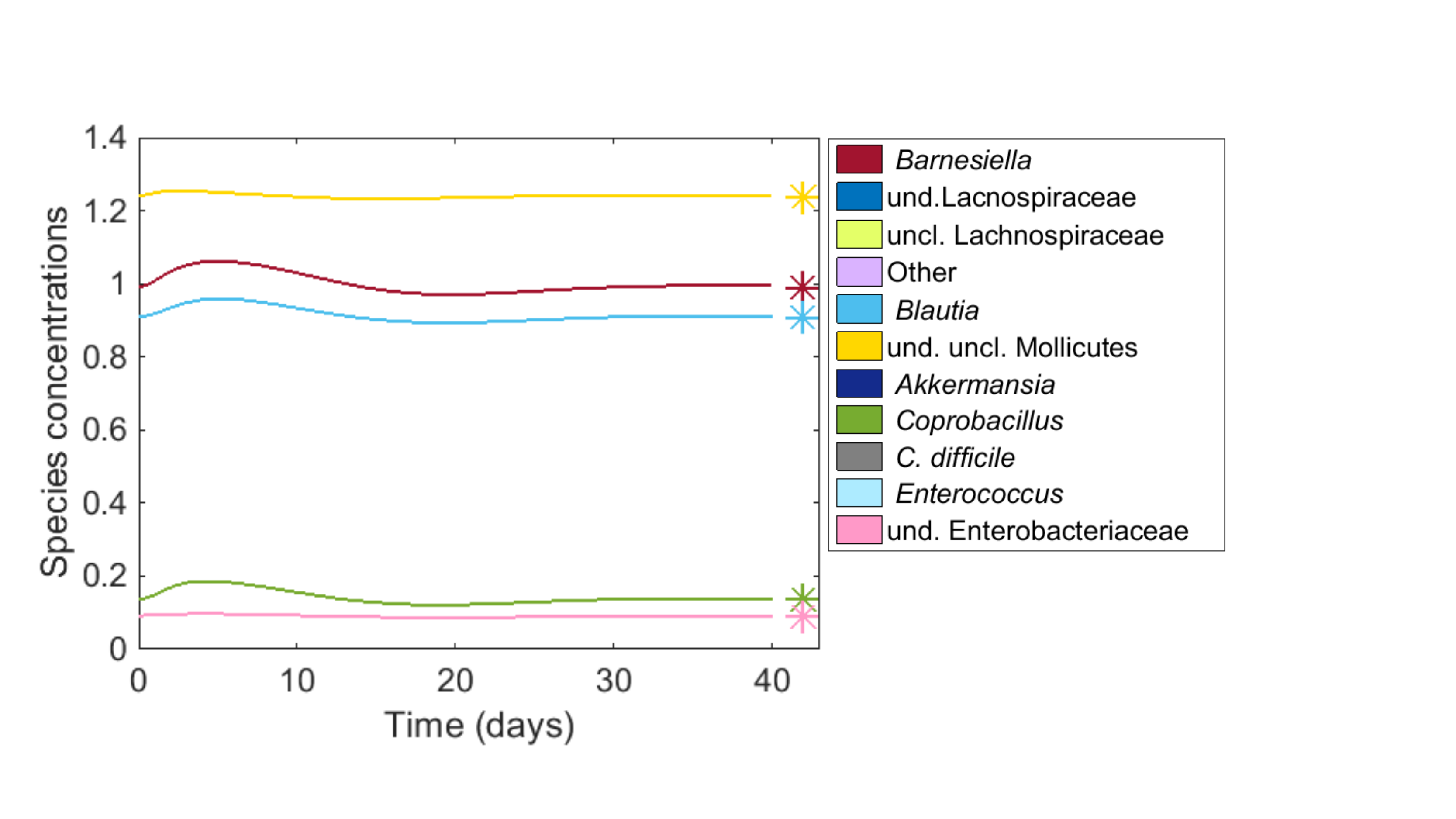}  
 \caption{Numerical integration of the controlled system (see equations~\eqref{added_node} in the main body). The equilibrium state stabilized upon injection of the controller (stars) is a slight modified version of the initially unstable fixed point, see table~\eqref{tab_s1}. The system is initialized out of equilibrium and, after a transient, converges to $\boldsymbol x^*$.}
   \label{fig_SI_1}
  \end{figure}

 \vspace{1 truecm}
  \textbf{\textit{Case B: Acting with one species of the pool to damp the concentration of the pathogens.}}
  Consider the stable fixed point $\boldsymbol x^*$ (table~\eqref{tab_s2}) composed by {\it Barnesiella}, und. Lachnospiraceae,  Other, {\it Blautia} and {\it C. difficile}. As explained in the paper, we now insert in the system another population, selected from the extended pool of interacting species. This latter configures as the controller. 
 Vector $\boldsymbol\alpha$ therefore follows from the interaction matrix $\boldsymbol A$. One can then calculate the fixed point that can be eventually attained by the system, given the specific selected controller.  
Retaining only the meaningful cases (fixed points with all positive entries), we obtain three possible solutions: $\boldsymbol x^*_L$ where the added species is uncl. Lachnospiraceae, $\boldsymbol x^*_M$ adding as external control the und. uncl. Mollicutes and $\boldsymbol x^*_E$ adding {und. Enterobacteriaceae} (see table~\ref{tab_s2}). From inspection of the obtained solutions, one can appreciate the impact of the different controllers employed: in the latter case the concentration of {\it C. difficile} stays almost constant, in the second example it increases, while in the first case it is reduced by a significant amount. 
 
   \begin{equation}
 \begin{tabular}{c | c | c | c | c}
  Populations & $\boldsymbol x^*$ & $\boldsymbol x^*_L$ & $\boldsymbol x^*_M$ & $\boldsymbol x^*_E$ \\ \hline 
  {\it Barnesiella} & 2.0745 & 2.5166 & 2.0480 & 2.2542\\
  und. Lachnospiraceae & 2.3607 & 1.9876 & 2.2674 & 2.7007\\
  Other & 1.9608 & 1.6929 & 1.9252 & 2.0842\\
  {\it Blautia} & 0.2724 & 0.1422 & 0.6171 & 0.1345\\
  {\it C. difficile} & 0.5402 & 0.2435 & 0.6194 & 0.5259\\
 \end{tabular}
 \stepcounter{equation}
\tag{S\theequation}
 \label{tab_s2}
 \end{equation}

   \begin{figure}
   \centering
   \includegraphics[width=12cm]{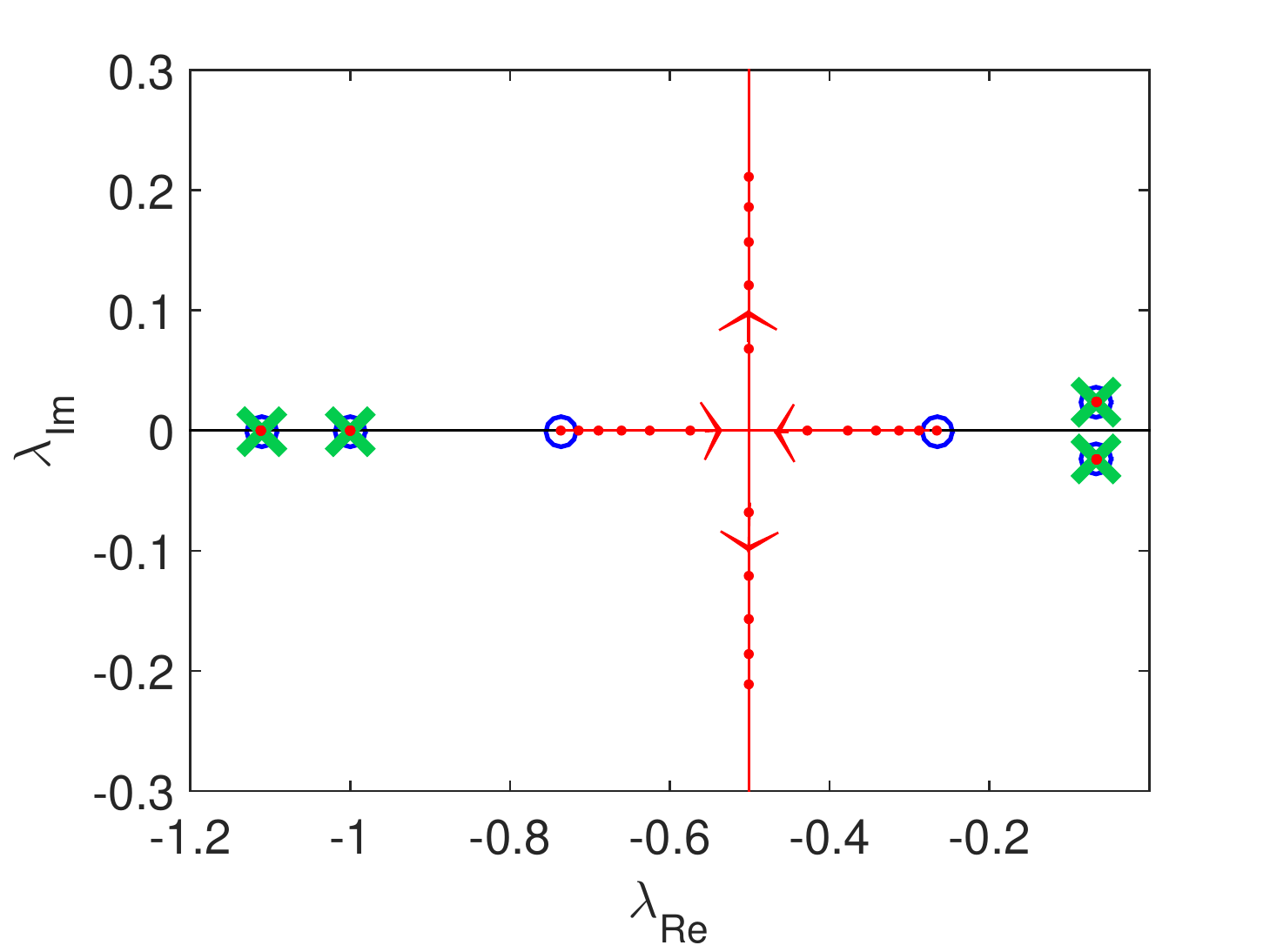}
   \caption{Root locus diagram relative to the Case B where, starting from the pool of 5 populations the control is performed adding the species of uncl. Lachnospiraceae. Blue circles correspond to the position in the complex plane of the roots of $\mathcal D(\lambda)$ (eigenvalues of $\boldsymbol J$ when $\rho=0$) while green crosses indicate the eigenvalues of the Jacobian in the limit $\rho\rightarrow\infty$. Paths followed by the eigenvalues when progressively increasing $\rho$ are shown by the red lines, while the red dots represent the solutions for discrete values of $\rho$, scanning the interval from $0$ to $0.1$.}
   \label{fig_SI_2}
  \end{figure}

  \vspace{1truecm}
  \textbf{\textit{Case C: Driving to extinction one species, the other being the target of the control.}}
   In this case we aim at enforcing the extinction of one of the species (here {\it C. difficile}), acting on the other ones. In other words the parameter vector $\boldsymbol\alpha$ (which characterizes the action of the control against the species) is forced to have the component relative to {\it C. difficile} equal to zero. At the same time, the corresponding entry of the vector $\boldsymbol x^*$ to be eventually stabilized is set to a negligible value. The concentrations of the species for respectively the initial and the final fixed points are compared in Figure~\ref{fig2} of the main body. The corresponding values of $\boldsymbol\alpha$ are also plotted. As a  supplementary material, we here report in Figure~\ref{fig_SI_3} the root locus diagram obtained for the case at hand.
   \begin{figure}
   \centering
   \includegraphics[width=12cm]{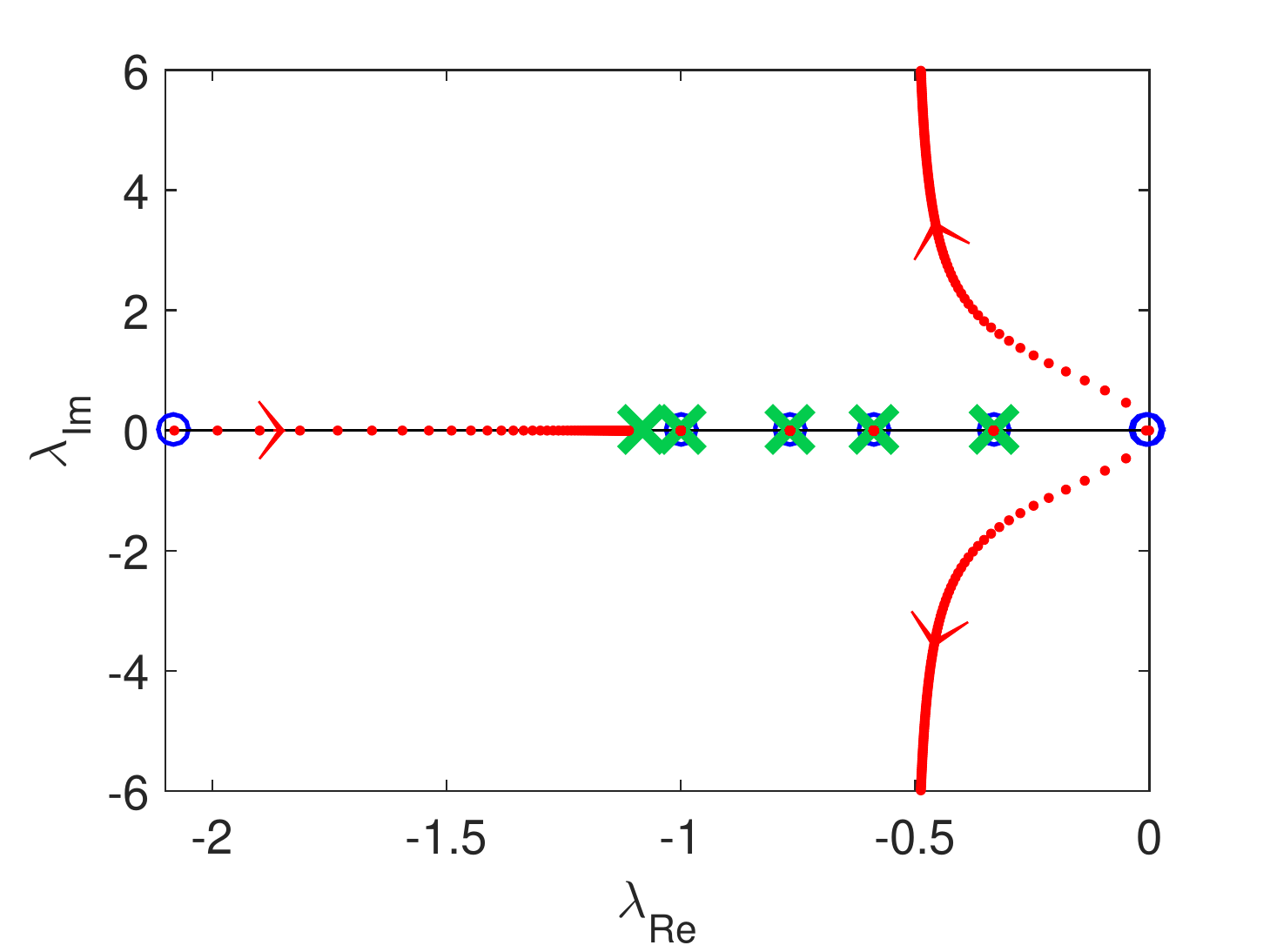}
   \caption{Root locus diagram relative to the Case C, where an indirect control is applied on the reduced system of $6$ species. For a description of the symbols, refer to the caption of Figure~\ref{fig_SI_2}.}
   \label{fig_SI_3}
  \end{figure}

 \vspace{1truecm}
 {\bf Exploiting a transient control to drive the system towards an existing stable fixed point}\\
 
  As anticipated in the main body of the paper, the control can be effectively employed  to steer the system towards a stable fixed point of the unperturbed dynamics, starting from out-of-equilibrium initial conditions. In this case, we set $\dot u = -\gamma u - \sum_j \beta_j (x_j-x_j^*)$. Here,  $x_j^*$ is an equilibrium solution of the uncontrolled dynamics, which proves linearly stable to external perturbations.
The parameter $\gamma$ can be tuned as desired so as to help the convergence towards $x_j^*$ without falling in the basin of attraction of other existing fixed points. As a proof of principle of the method, we choose a stable fixed point of the global macrobiota ecosystem, i.e. including the complete pool of $11$ populations. This is characterized by $x_1^*=9.299$,  $x_3^*=12.3085$, $x_4^*=3.1627$ and $x_j^*=0$ for $j \ne 1,3,4$. The largest real part of the eigenvalues of the associated $11 \times 11$ Jacobian matrix turns out to be $(\lambda_{Re}) =-0.1306<0$, thus implying stability of the aforementioned equilibrium. Imagine to initialize the system out of equilibrium with all species, including the pathogen {\it C. difficile}, being assigned a random concentration $x_j(0) \ne 0$. The system is let evolve for a while and then, at time $t^*$, the control is injected. Here, ${\boldsymbol \alpha}$ and ${\boldsymbol \beta}$ are assigned as random, uniformly distributed over [0,1], parameters. As clearly displayed in Figure~\ref{fig_SI_4}, the system is steadily moved towards the equilibrium $x_j^*$ (stars), while the control converges to zero after an abrupt jump. In other words, after a transient whose duration depends on the chosen parameters, the system achieves its asymptotic (pathogen free) equilibrium and the control can be safely disconnected.
 
\begin{figure}
 \centering
 \subfigure[]
   {\includegraphics[width=10cm]{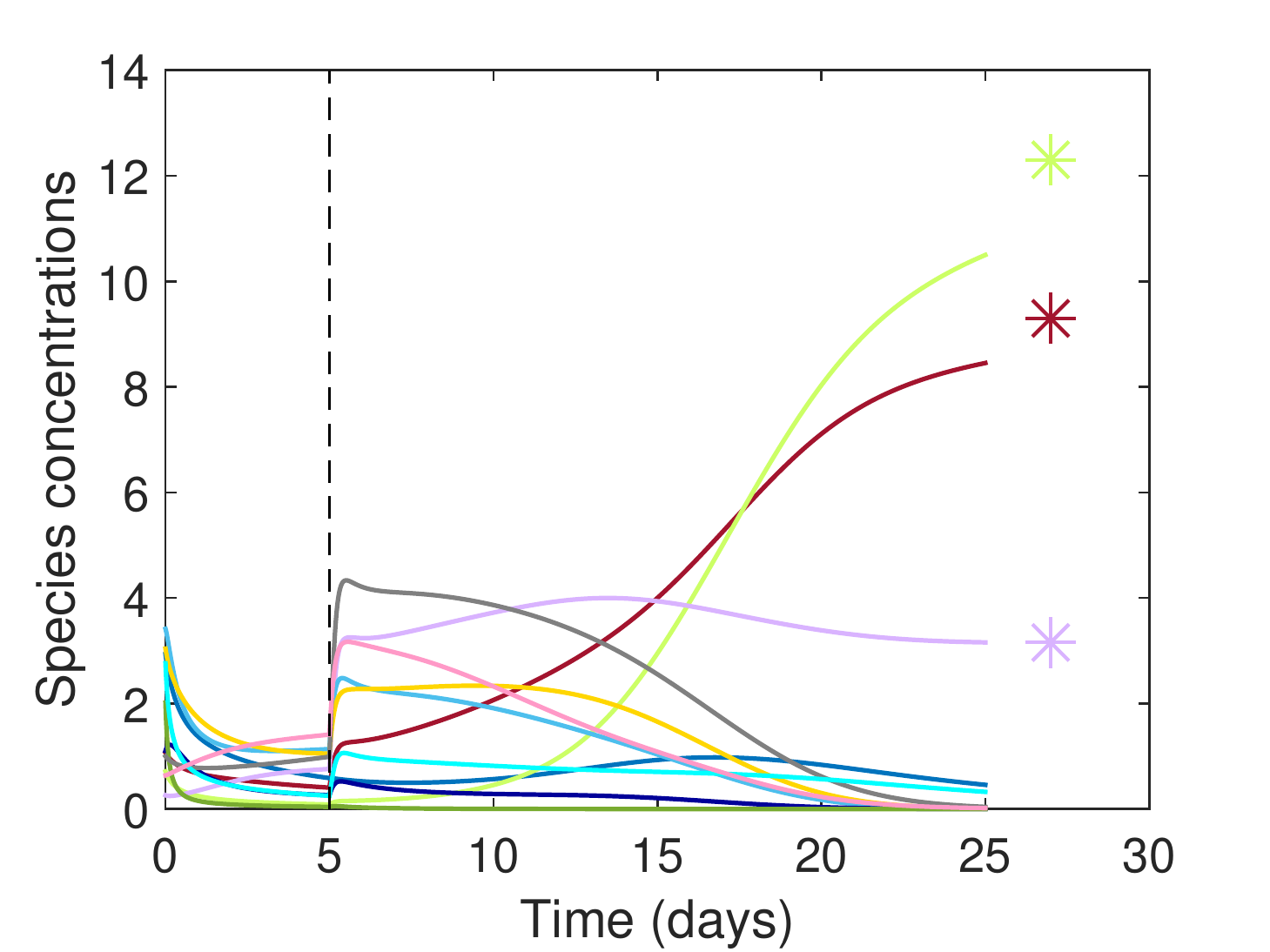}}
 \hspace{5mm}
 \centering
 \subfigure[]
   {\includegraphics[width=10cm]{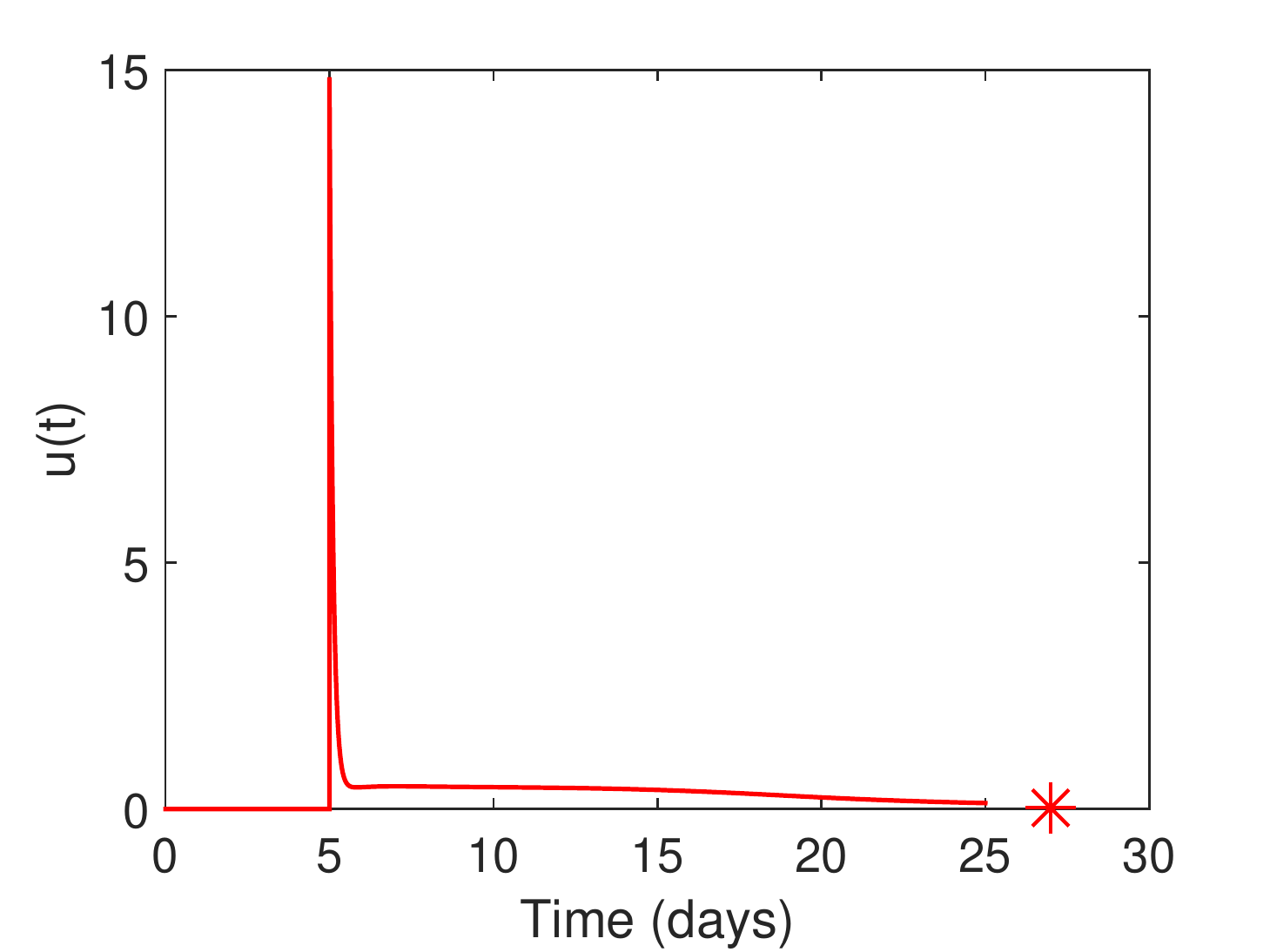}}
 \hspace{5mm}
  \caption{Driving the system towards a stable fixed point of the unperturbed dynamics. The system is initiated out-of-equilibrium: the concentration of all species, including the pathogen {\it C. difficile}, is set to values different from zero. At $t^*=5$ days the control $u$ is injected. After a transient the system converges to its stable equilibrium characterized by  $x_1^*=9.299$,  $x_3^*=12.3085$, $x_4^*=3.1627$ and $x_j^*=0$ for $j \ne 1,3,4$, while the control $u$ is turned to zero. Here $\gamma=10$ and $u(0)=15$.}
   \label{fig_SI_4}
  \end{figure}

 \vspace{1truecm}
{\bf On the controllability condition}\\
  
Observe that the control procedure here discussed requires computing the vector $\boldsymbol\beta$. Determining this latter implies inverting a matrix, an operation that imposes a mathematical constraint that we shall hereafter analyze more in depth. The matrix to be inverted, as defined in the main text, reads:
   \begin{equation}
  H_{nm}=\sum_{k=0}^{N-n}c_{k+n+1}(\boldsymbol{G}^k\boldsymbol{q})_m
  \stepcounter{equation}
\tag{S\theequation}
  \label{H}
 \end{equation}
 where $c$ stands for the coefficients\footnote{\label{foot}
 \begin{equation}
 \begin{split}
  &c_1=\det(\boldsymbol G)\\
  &c_k=\frac{(-1)^{k-1}}{(k-1)!}\sum_{\substack{j_1...j_{k-1}=1\\j_1\neq...\neq j_{k-1}}}^N \det\boldsymbol G_{(j_1j_1)(j_2j_2)\dots(j_{k-1}j_{k-1})} \text{  for } k=2,\dots,N-1\\
  &c_N=(-1)^{N-1}Tr(\boldsymbol G)\\
  &c_{N+1}=(-1)^N
  \end{split}
  \label{coeff_c}
  \stepcounter{equation}
\tag{S\theequation}
 \end{equation}
 where with $\boldsymbol G_{(j_1j_1)(j_2j_2)\dots(j_{k-1}j_{k-1})}$ we identify the minor obtained from matrix $\boldsymbol G$ by removing the $j_1$-th, $j_2$-th,...,$j_{k-1}$-th rows and columns.
} of the characteristic polynomial of $\boldsymbol G$, namely
 $\det(\boldsymbol{G}-\lambda\boldsymbol{I})=\sum_{l=0}^{N}c_{l+1}\lambda^l$. The invertibility is ensured if $\det(\boldsymbol H)\neq0$, in formulae:
 \begin{equation}
 \begin{split}
  \det(\boldsymbol H)&=\sum_{i_1=1}^N \dots \sum_{i_N=1}^N \epsilon_{i_1,...i_N} H_{1i_1}H_{2i_2} \dots H_{N-1i_{N-1}}H_{Ni_N}=\\
  &=\sum_{i_1=1}^N \dots \sum_{i_N=1}^N \epsilon_{i_1,...i_N}\Biggl[\sum_{k_1=0}^{N-1}c_{k_1+2}(\boldsymbol{G}^{k_1}\boldsymbol{q})_{i_1}\Biggr]\Biggl[\sum_{k_2=0}^{N-2}c_{k_2+3}(\boldsymbol{G}^{k_2}\boldsymbol{q})_{i_2}\Biggr]\dots\\
  &\dots\Biggl[\sum_{k_{N-1}=0}^{1}c_{k_{N-1}+N}(\boldsymbol{G}^{k_{N-1}}\boldsymbol{q})_{i_{N-1}}\Biggr][c_{N+1}q_{i_N}] \ne 0
  \end{split}
  \stepcounter{equation}
\tag{S\theequation}
 \end{equation}
 where $\epsilon_{i_1,...i_N}$ is the Levi-Civita tensor. This complicated expression can be heavily simplified. The Levi-Civita symbol is in fact totally antisymmetric in the permutation of its indices. As a consequence,  all the terms multiplied by $\epsilon_{i_1,...i_N}$ which are symmetric in the permutations, cancel out. It follows that the terms containing the product of two or more factors $(\boldsymbol{G}^{k}\boldsymbol{q})$ with the same power $k$ are identically equal to zero. The only terms which surviveare those obtained by just retaining the largest possible value of$k$ in each summation ($k_1=N-1$, $k_2=N-2$,..., $k_{N-1}=1$). In formulae:
 \begin{equation}
 \begin{split}
  \det(\boldsymbol H)&=\sum_{i_1=1}^N \dots \sum_{i_N=1}^N \epsilon_{i_1,...i_N}c_{N+1}(\boldsymbol{G}^{N-1}\boldsymbol{q})_{i_1}c_{N+1}(\boldsymbol{G}^{N-2}\boldsymbol{q})_{i_2}\dots c_{1+N}(\boldsymbol{G}\boldsymbol{q})_{i_{N-1}}c_{N+1}q_{i_N}=\\
  &=(-1)^{N^2}\sum_{i_1=1}^N \dots \sum_{i_N=1}^N \epsilon_{i_1,...i_N}(\boldsymbol{G}^{N-1}\boldsymbol{q})_{i_1}(\boldsymbol{G}^{N-2}\boldsymbol{q})_{i_2}\dots(\boldsymbol{G}\boldsymbol{q})_{i_{N-1}}q_{i_N}.
  \end{split}
  \stepcounter{equation}
\tag{S\theequation}
  \label{detH}
 \end{equation}
 where use has been made of the fact that $c_{N+1}=(-1)^N$ (see equation~\eqref{coeff_c} in the footnote~\ref{foot}).\\
 
 Drawing on this preliminary observations it is possible to re-interpret the above controllability constraint, making contact with standard control theory. The controllability condition amounts to require that the matrix 
 \begin{equation}
  \boldsymbol C\equiv[\boldsymbol q,\boldsymbol{Gq},...,\boldsymbol{G}^{N-1}\boldsymbol{q}]=
 \left (
  \begin{matrix}
  q_1  &  (\boldsymbol{Gq})_1  &  (\boldsymbol{G}^2\boldsymbol{q})_1  &  \hdots  &  (\boldsymbol{G}^{N-1}\boldsymbol{q})_1\\
  q_2  &  (\boldsymbol{Gq})_2  &  (\boldsymbol{G}^2\boldsymbol{q})_2  &  \hdots  &  (\boldsymbol{G}^{N-1}\boldsymbol{q})_2\\
  \vdots\\
  q_N  &  (\boldsymbol{Gq})_N  &  (\boldsymbol{G}^2\boldsymbol{q})_N  &  \hdots  &  (\boldsymbol{G}^{N-1}\boldsymbol{q})_N\\\\
 \end{matrix}
 \right )
 \stepcounter{equation}
\tag{S\theequation}
 \label{M}
 \end{equation}
 has maximum rank. Here, with the notation $(\boldsymbol{G}^l\boldsymbol{q})_k$ we identify the $k$-th entry of the vector obtained from the product of the matrix $\boldsymbol G$ to the power of $l$ with vector $\boldsymbol q$. In system theory the matrix $\boldsymbol C$ is called the {\it controllability} matrix of the pair $(\boldsymbol G,\boldsymbol q)$. Since $\boldsymbol C$ is a square matrix, the maximum rank condition is equivalent to require $\det(\boldsymbol C)\neq0$, in formulae:
 \begin{equation}
 \begin{split}
  \det(\boldsymbol C)&=\sum_{\sigma_1=1}^N \dots \sum_{\sigma_N=1}^N \epsilon_{\sigma_1,...,\sigma_N}C_{\sigma_11}C_{\sigma_22}\dots C_{\sigma_{N-1}N-1}C_{\sigma_NN}=\\
  &=\sum_{\sigma_1=1}^N \dots \sum_{\sigma_N=1}^N \epsilon_{\sigma_1,...,\sigma_N}(\boldsymbol{G}^{0}\boldsymbol{q})_{\sigma_1}(\boldsymbol{G}^{1}\boldsymbol{q})_{\sigma_2} \dots (\boldsymbol{G}^{N-2}\boldsymbol{q})_{\sigma_{N-1}}(\boldsymbol{G}^{N-1}\boldsymbol{q})_{\sigma_N}=\\
  &=(-1)^N\sum_{\sigma_1=1}^N \dots \sum_{\sigma_N=1}^N \epsilon_{\sigma_1,...,\sigma_N}(\boldsymbol{G}^{N-1}\boldsymbol{q})_{\sigma_1}(\boldsymbol{G}^{N-2}\boldsymbol{q})_{\sigma_2} \dots (\boldsymbol{G}\boldsymbol{q})_{\sigma_{N-1}}(\boldsymbol{q})_{\sigma_N} \ne 0
  \end{split}
  \stepcounter{equation}
\tag{S\theequation}
  \label{detM}
 \end{equation}
 where use has been made of the definition of $\boldsymbol C$~\eqref{M}, namely $C_{ij}=(\boldsymbol{G}^{j-1}\boldsymbol{q})_i$.\\
 Expressions~\eqref{detH} and~\eqref{detM} are identical (except for the sign) and consequently the two conditions, $\det(\boldsymbol H)\neq0$ e $\det(\boldsymbol C)\neq0$, prove equivalent. Stated differently,  the condition of invertibility of matrix $\boldsymbol H$, obtained as a self-consistent constraint for the introduced control scheme, coincides with
 the standard controllability condition, as known in control theory.\\
 As a final remark we recall (see main body) that our goal is not to arbitrarily assign the polynomial $\mathcal N(\lambda)$ but rather to locate its roots $z_k$ within the open left-hand plane. In this respect, the necessary and sufficient system-theoretic condition is the so-called {\it stabilizability} of the pair $(\boldsymbol G, \boldsymbol q)$, which results in the Popov-Belevitch-Hautus rank condition~\cite{Kailah80, Yuan_etal13}.

 \end{section}

\addcontentsline{toc}{chapter}{Bibliography}
\bibliography{myBib}{}
\bibliographystyle{ieeetr}

\end{document}